## Galaxy Scale Interstellar Infrared Spectrum Reproduced By
## A Hydrocarbon Pentagon-Hexagon Combined Molecule

NORIO OTA

Graduate School of Pure and Applied Sciences, University of Tsukuba,

1-1-1 Tenoudai Tsukuba-city 305-8571, Japan;   n-otajitaku@nifty.com

Interstellar dust shows ubiquitous interstellar infrared spectrum (IR) due to polycyclic aromatic hydrocarbon (PAH). By our previous quantum chemistry calculation, it was suggested that a molecule group having hydrocarbon pentagon-hexagon combined skeleton could reproduce observed IR of dust clouds in Milky Way galaxy. This paper extends to other many galaxies. Typical galaxies are NGC6946 and M83. Those infrared spectrum were compared with that of a model molecule (C23H12)2+ having hydrocarbon two pentagons combined with five hexagons. Observed major infrared bands of 6.2, 7.7, 8.6, and 11.3 micrometer were successfully reproduced as 6.4, 7.7, 8.5, and 11.2 micrometer. Even observed weaker bands of 12.0, 12.7, 14.2 micrometer were predicted well by computed bands as 12.0, 12.6, and 13.9 micrometer. IR intensity ratio was compared to check theoretical validity.  Calculated intensity ratio between 7.7 versus 11.3 micrometer (PAH7.7/11.3) was 4.0, whereas observed ratio was in a range of 2~6, also calculated PAH6.2/11.3 was 1.4 for observed range of 0.9~2.6. Especially, every calculated ratio was so close to that of M83 arm region. It should be noted that both calculated wavelength and intensity could reproduce observed galaxy scale infrared spectrum. Hydrocarbon pentagon-hexagon molecule would be general carrier in many galaxies including Milky Way.

## 1, INTRODUCTION

Interstellar infrared spectrum (IR) due to polycyclic aromatic hydrocarbon (PAH) was ubiquitously observed in our Milky Way galaxy from 3 to 20µm (Boersma et al. 2013, 2014). However, any single PAH molecule or related species showing universal infrared spectrum has not yet been identified to date. Identification is essentially important to search chemical evolution step of organics and to study material building block of creation of life in the universe (Ota 2016). In 2014, accompanying a material study of void induced graphene sheet (Ota 2014a), it was incidentally founded that void induced graphene molecule $(C_{23}H_{12})^{2+}$ shows very similar infrared spectrum with interstellar observed one (Ota 2014b, 2015a). This molecule contains two hydrocarbon pentagons combined with five hexagons. This may lead to an identification of specific carrier molecule for interstellar infrared emission. After that, many PAH molecules were test for identification. Simple molecule $(C_{12}H_8)^{3+}$ had also show good coincidence with observed strong bands, which configuration was hydrocarbon one pentagon combined with two hexagons (Ota 2015b).  Recently, based on observation data on nebula NGC 2023 (Peeters 2017), we could find out that among 16 observed bands, 14 bands were successfully reproduced well by calculation (Ota 2017b). Also, I traced a history of star's death and birth and made one possible hypothesis named "top down hydrocarbon evolution theory" (Ota 2017a), that is, nucleation and diffusion of graphene molecule after supernova expansion, void creation by high speed proton, hydrogenation by low speed proton, and ionization by high energy photon. Energy diagram of each modification step of $(C_{23}H_{12})^{2+}$ was calculated. We could estimate the central star mass to be 4~7 times heavier than our sun.

Above observation and coincidence were in our Milky Way galaxy. Question is how about a situation on other galaxy. This paper extends such an approach to many galaxies. There are several galaxy scale infrared spectrum observation as like NGC6946 (Sakon 2007), M83 (Vogler 2005), and systematic data on 129 galaxies (Leja 2016). Those observed infrared spectrum should be compared in detail with that of a model molecule $(C_{23}H_{12})^{2+}$. It is so amazing to compare their sizes, one is $10^{20}$ meter another is $10^{-9}$ meter. Size difference is $10^{29}$ times. In this paper, we could confirm good coincidence of infrared spectrum both on wavelength and intensity. Hydrocarbon pentagon-hexagon combined molecule will be ubiquitous in the great universe.

## 2, MODEL MOLECULES AND CALCULATION METHOD

In this study, a specific molecule $(C_{23}H_{12})^{2+}$ having hydrocarbon two pentagons and five hexagons was modeled as a typical interstellar infrared spectrum carrier molecule. In calculation, we have to obtain total energy, optimized atom configuration, and infrared vibrational mode frequency and strength depend on a given initial atomic configuration, charge and spin state Sz. Density functional theory (DFT) with unrestricted B3LYP functional was applied utilizing Gaussian09 package (Frisch et al. 1984, 2009) employing an atomic orbital 6-31G basis set. The first step calculation is to obtain the self-consistent energy, optimized atomic configuration and spin density. Required convergence on the root mean square density matrix was less than $10^{-8}$ within 128 cycles. Based on such optimized results, harmonic vibrational frequency and strength was calculated. Vibration strength is obtained as molar absorption coefficient ε (km/mol.). Comparing DFT harmonic wavenumber $N_{DFT}$ (cm$^{-1}$) with experimental data, a single scale factor 0.965 was used (Ota 2015b). Concerning a redshift for the anharmonic correction, in this paper we did not apply any correction to avoid over estimation in a wide wavelength representation from 2 to 30 micrometer.

Corrected wave number N is obtained simply by N (cm$^{-1}$) = $N_{DFT}$ (cm$^{-1}$) x 0.965.

Wavelength λ is obtained by λ (micrometer) = 10000/N(cm$^{-1}$).

Reproduced IR spectrum was illustrated in a figure by a decomposed Gaussian profile with full width at half maximum FWHM=4cm$^{-1}$.

## 3, INFRARED SPECTRUM OF STAR BURST GALAXY NGC6946

NGC6946 is 22 million light-years away spiral galaxy including active star burst area and many supernovas (9 in last 100 years). Detailed infrared observation was down by Itsuki Sakon et al., (Sakon 2007) using AKARI infrared survey. In Figure 1 top, infrared spectrum of arm region is illustrated. Major bands are 6.2, 7.6, 7.8, 8.6 and 11.3 micrometer. Inter-arm region also shows similar major bands. Those are similar with ubiquitously observed bands in Milky Way's dust clouds. Quantum chemically calculated infrared spectrum of $(C_{23}H_{12})^{2+}$ was illustrated in a down part of Figure 1. Resulted major bands were 6.4, 7.4, 7.7, 8.2, 8.5, and 11.2 micrometer. We could fairly nice coincidence between galaxy scale observation and a single molecule calculation.

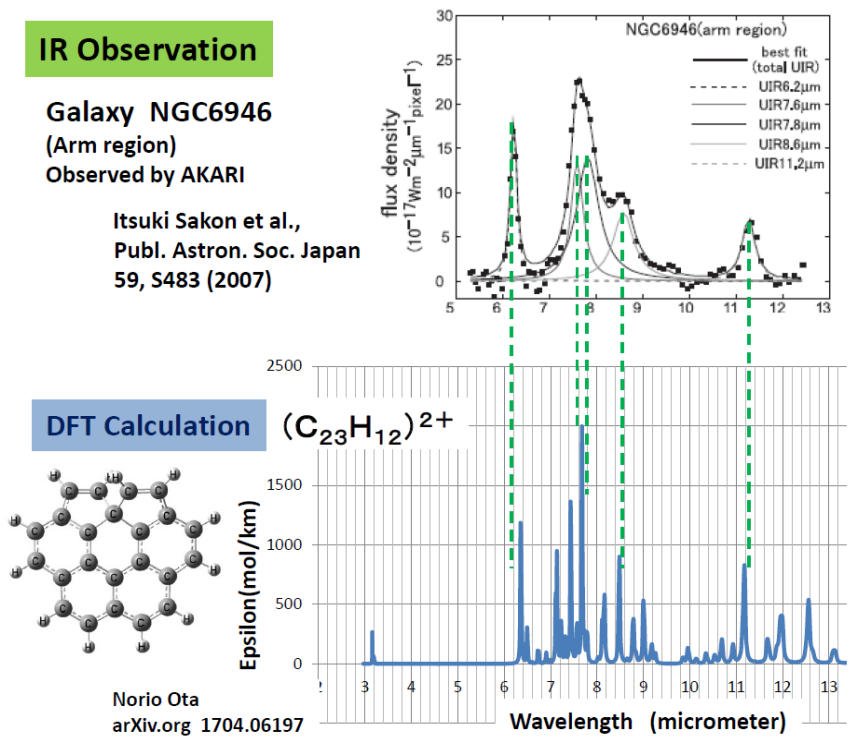

Figure 1, Comparison of observed infrared spectrum of NGC6946 arm region by AKARI and a quantum chemistry calculated spectrum of $(C_{23}H_{12})^{2+}$, which show fairly nice coincidence.

4, INFRARED SPECTRUM OF SUPERNOVA ACTIVE GALAXY M83

 Galaxy M83 is 15 million light-years away barred spiral galaxy. It is supernova active galaxy, of which six supernovas were discovered after 1923. Infrared spectrum was surveyed by A. Vogler et al. (Vogler 2005) by ISOCAM as shown on top of Figure 2, D e t a i l e d five area's infrared spectrum were noted by (1)~(5).  They show common major spectrum at 6.2, 7.7, 8.6, and 11.3 micrometer marked by green dotted lines. Calculated IR of $(C_{23}H_{12})^{2+}$ was compared on down part, of which major bands were 6.4, 7.7, 8.5, and 11.2 micrometer. Also weaker observed bands were marked by blue dotted lines at 8.2, 12.0, 12.7, 13.6 and 14.2 micrometer. Again, calculated related bands were 8.2, 12.0, 12.6, 13.6, and 14.0 micrometer. Calculated 7.1 and 9.0 micrometer bands are accidentally overlapped with Ar II and Ar III atomic emission lines marked by light blue. They may be not noticed as PAH original bands. Totally, we could reproduce very details of M83 spectrum by a single model molecule beyond $10^{29}$ size difference and 15 million light years away distance.

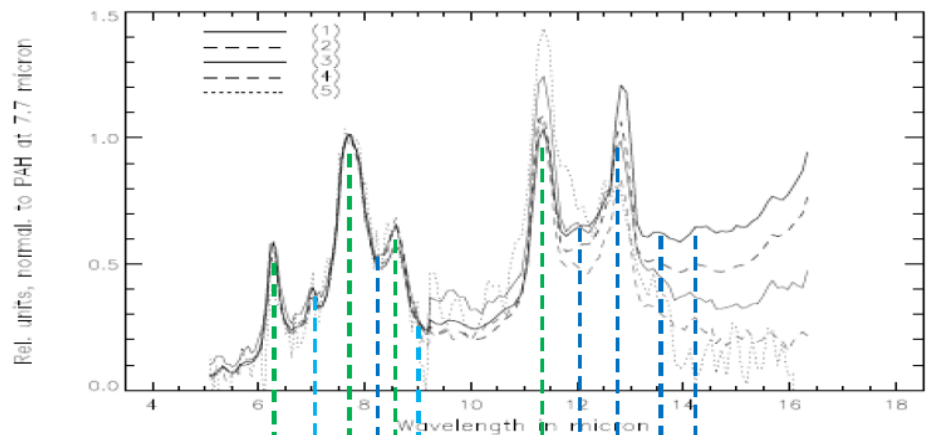

**IR Observation, M83  Galaxy**

Vogler, A.  et al.,
A&A, 441,491 (2005)

**DFT Calculation**
$(C_{23}H_{12})^{2+}$

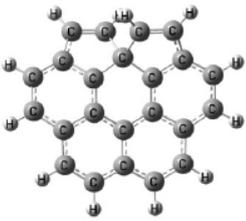

Norio Ota
arXiv.org  1704.06197

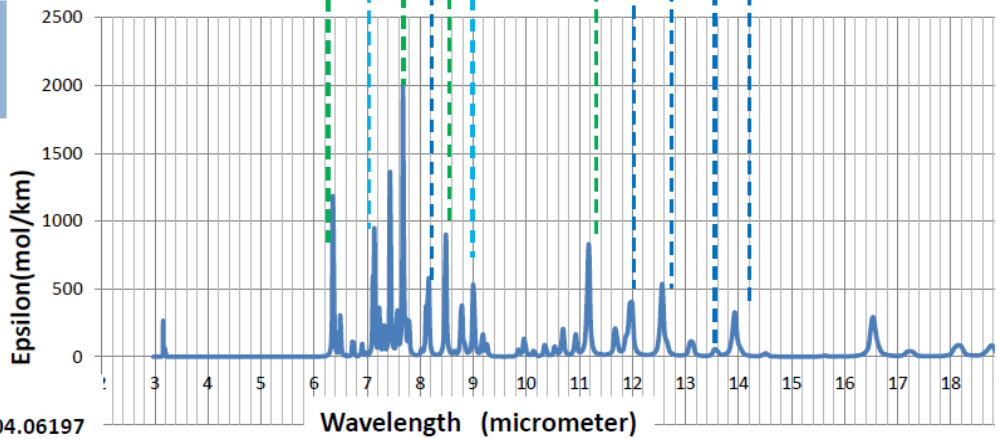

Figure 2, Observed infrared spectrum of galaxy M83 (Vogler 2005) compared with calculated bands of a single model molecule $(C_{23}H_{12})^{2+}$ having hydrocarbon two pentagons combined with five hexagons. Green dotted lines are major observed bands, blue are weaker bands and light blue are accidentally overlapped bands with Ar II and Ar III. Calculated bands could reproduce detailed feature of M83.

## 5, INFRARED SPECTRUM OF 129 GALAXIES

Broadband photometry of 129 galaxies was studied by J. Leja et al. (Leja 2017). Among many important observed data, mid infrared spectrum was classified by a specific parameter $Q_{PAH}$, which is the fraction of total dust mass in PAHs. Upper illustration in Figure 3 is observed result showing featured PAH spectrum. Observed major wavelengths in log scale are 6.2, 7.7, 8.6, and 11.3 micrometer (green dotted lines), weaker one are 7.1, 12.0, 12.7, 14.2 micrometer (blue dotted lines). Again, as shown in a down part of Figure 3, calculated wavelength were 6.4, 7.7, 8.5, 11.2, 7.1, 12.0, 12.6, and 14.0 micrometer respectively. We could easily understand good coincidence of infrared spectrum between galaxy scale observation and molecular scale calculation.

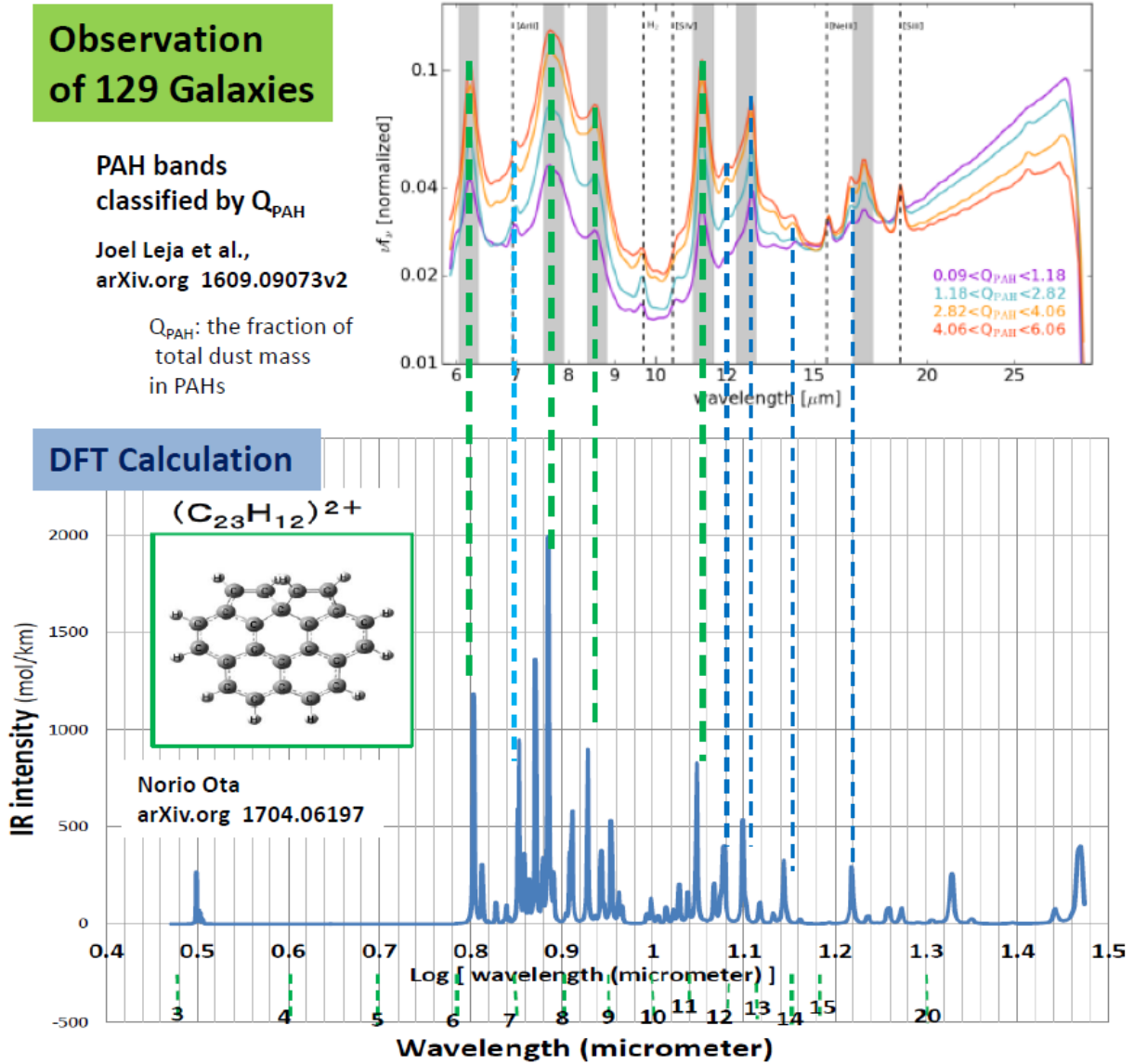

Figure 3, Observed mid infrared spectrum of 129 galaxies were edited by Joel Leja et al. (Leja 2016) as shown on upper illustration, which were classified by parameter $Q_{PAH}$ as the fraction of total dust mass in PAHs. Such observed spectrum feature could be reproduced very well by the quantum chemical calculation of a single molecule $(C_{23}H_{12})^{2+}$ as illustrated on down.

## 6, INFRARED SPECTRUM INTENSITY RATIO

It is important to study IR intensity ratio to certify theoretical validity.  M. Yamagishi et al. opened interesting data on IR intensity of interstellar dust cloud of M17SW in Milky Way as shown in Figure 4, where many small black dots show spatial variation of  IR intensity ratio (Yamagishi 2016). For example, PAH7.7 denotes IR intensity of 7.7 micrometer band, and PAH7.7/PAH11.3 show intensity ratio of 7.7 micrometer band versus 11.3 micrometer band. On these basic illustrations, we overwrite galaxy scale IR intensity ratio utilizing data edited by I. Sakon et al.  (Sakon 2007), which are M83 arm region by blue triangle, M83 interarm by red triangle, NGC6946 arm by blue circle, NGC6946 interarm by red circle. Whereas, we estimated IR intensity of molecule $(C_{23}H_{12})^{2+}$ by assuming that intensity is proportional to peak height because of its narrow band width of FWHM=4cm$^{-1}$. As noted in Table 1, major intensity ratio was obtained, where calculated PAH7.7 was a sum of 7.43 and 7.68 micrometer peaks. Resulted major intensity ratio ware as follows,

> PAH7.7/PAH11.3=4.04
> PAH6.2/PAH11.3=1.42
> PAH8.6/PAH11.3=1.08
> PAH12.7/PAH11.3=0.65

In Figure 4, calculated value was marked by red oval. In an illustration (A), we can see lineally increasing PAH6.2/PAH11.3 with increasing PAH7.7/PAH11.3. Galaxy scale observed points sit on a same line of M17SW of Milky Way. Calculated point (red oval mark) of $(C_{23}H_{12})^{2+}$ was so close to observed M83 arm region (blue triangle) and NGC6946 arm region (blue circle), but somewhat far from interarm region (M83 by red triangle and NGC6946 by red circle). Similar tendency was also discovered in (B), PAH8.6/PAH11.3 increases with PAH7.7/PAH11.3. Calculated value was very close to M83 arm region data, but far from interarm one. Additional calculated ratio between PAH12.7/PAH11.3 versus PAH7.7/PAH11.3 was marked in (C).  Calculated value overlapped with data points of Milky Way M17SW. Unfortunately, there is no observed data on galaxies.

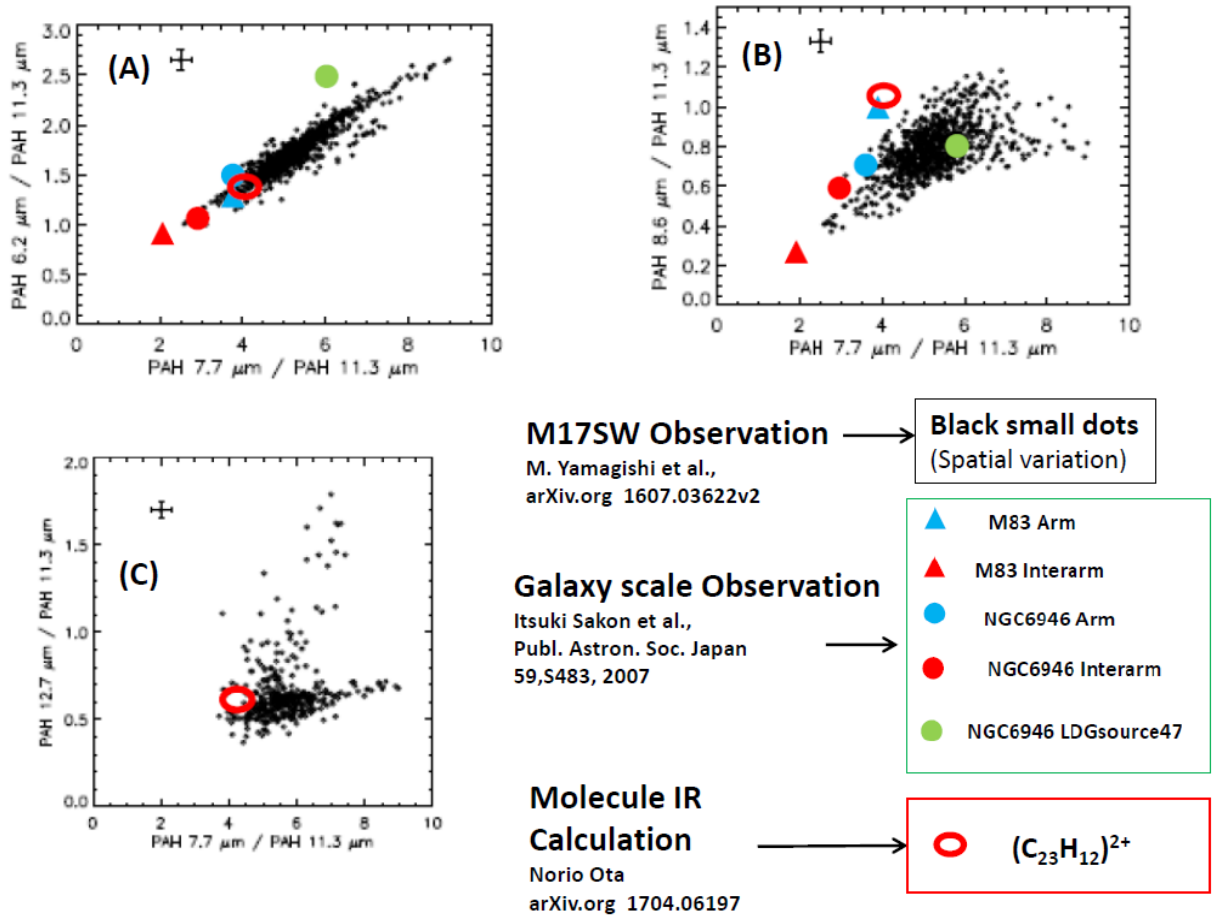

Figure 4, Infrared spectrum intensity ratio of calculated values of $(C_{23}H_{12})^{2+}$ (red oval) compared with observed galaxy scale data. Calculated value was so close with observed value of M83 arm region (blue triangle)

Table 1, Calculated infrared intensity and intensity ratio of $(C_{23}H_{12})^{2+}$.

| Band name | Wavelength (micrometer) | Intensity ( mol/km) |
|---|---|---|
| PAH6.2 | 6.35 | 1172 |
| PAH7.7 | 7.43 | 1339 |
| PAH7.7 | 7.68 | 1995 |
| PAH8.6 | 8.49 | 890 |
| PAH11.3 | 11.18 | 825 |
| PAH12.7 | 12.56 | 537 |
| | | |
| Intensity Ratio | PAH7.7/PAH11.3=(1339+1995)/825=4.04 | |
| | PAH6.2/PAH11.3=1172/825=1.42 | |
| | PAH12.7/PAH11.3=537/825=0.65 | |
| | PAH8.6/PAH11.3=890/825=1.08 | |

## 6, CONCLUSION

Infrared spectrum of a model molecule $(C_{23}H_{12})^{2+}$ having hydrocarbon two pentagons combined with five hexagons was obtained by a quantum chemistry based DFT calculation, which was compared with galaxy scale observed infrared spectrum. Typical galaxies are NGC6946 and M83.

(1) Observed strong infrared bands of 6.2, 7.7, 8.6, and 11.3 micrometer were successfully reproduced as 6.4, 7.7, 8.5, and 11.2 micrometer.

(2) Weaker bands of 12.0, 12.7, 14.2 micrometer were predicted well by computed bands as 12.0, 12.6, and 13.9 micrometer. Calculated 7.1 and 9.0 micrometer bands are accidentally overlapped with ArII and ArIII atomic emission.

(3) IR intensity ratio was compared to check theoretical validity. Calculated PAH7.7/11.3 band ratio was 4.0, whereas observed ratio were 2~6, also calculated PAH6.2/11.3 was 1.4 for observed range 0.9~2.6. Model molecule intensity ratio was very close to that of M83 arm region.

It should be noted that both calculated wavelength and intensity could reproduce observed galaxy scale infrared spectrum. This paper suggests that hydrocarbon pentagon-hexagon molecule may be ubiquitous in many galaxies including Milky Way.